\documentclass[final,3p,times,twocolumn]{elsarticle}
\usepackage{graphicx}
\usepackage{lineno}
\usepackage[table,xcdraw]{xcolor}
\journal{Carbon}
\begin{document}
\begin{frontmatter}

\title{Process Yield and Device Stability Improvement for Sol-gel Alumina Passivation layer based GFETs}

 \author[label*]{Nama Premsai}

\address{Department of Electrical engineering}
\address{Indian Institute of Technology Bombay Mumbai-76, India}
\address{email: premsainama@ee.iitb.ac.in, premsaii45@gmail.com}
\begin{abstract}

The stability of GFET devices is a major problem and needs a good passivation layer for the same. Low cost and low-temperature sol-gel alumina passivation layer-based GFETs is studied here with the goals to improve stability of the device and achieve high process yield. The  process yield and device stability are explored for two different molarities of 0.1 M and 0.05 M. The parameters like mobility, trap charge density, minimum conductivity plateau width and minimum conductivity are extracted to compare the stability of devices. The results indicate that GFETs with 0.1 M have problems of process yield due to crack formation in the channel region post-annealing, where close to half of the devices are working, and also working devices are  are not stable and degrading very fast within a week's time. On the other hand, 0.05 M sol-gel Alumina-based sample exhibits 100$\%$ process yield with all working devices and observed stable behavior for more than two months. Hence we propose an optimized process recipe for a sol-gel Alumina-based passivation layer to achieve the best possible process yield and device stability for GFETs.

\end{abstract}

\begin{keyword}
Monolayer graphene \sep  GFET \sep sol-gel Alumina \sep Dirac point \sep DUV annealing

\end{keyword}

\end{frontmatter}

\section{Introduction}
\label{S:1}

The ambipolar carrier modulation through the electric field \cite{novoselov2004electric}, the absence of bandgap and the remarkable electronic properties like high carrier mobility \cite{bolotin2008ultrahigh} with ballistic transport of the carriers \cite{du2008approaching} and high-velocity saturation \cite{shishir2009velocity} have convinced the device community to explore the two-dimensional material graphene for the high-speed radio frequency (RF) applications \cite{han2014graphene}. Along with the exceptional electronic properties, CMOS process compatibility, mechanical strength, and flexibility \cite{lee2012multi,park2016extremely} have made graphene a popular material for transistor devices. 

Besides several advantages of graphene transistors, there are few challenges like yield problem and unintentional doping which contributes to the total stability of GFETs. The yield has been contributed from the several factors in particular (i) defect free CVD graphene growth \cite{ullah2021graphene} (ii) careful graphene transfer (delamination from growth substrate, handling during transfer and removal of the handling material) \cite{ullah2021graphene} (iii) good dielectric environment such as substrate surface \cite{hu2014substrate}, passivation \cite{bzurutuzaaelorza2015highly,alexander2016encapsulation} and top gate dielectric \cite{chen2009dielectric, liao2010graphene,wang2018high} (iv) proper selection of contacts \cite{giubileo2017role}. The process stability in each fabrication step add to the total yield of the device \cite{neumaier2019integrating}. 

The unintentional doping in graphene transistors are due to the environment adsorbates, residues from the fabrication processes that limits the usage of the graphene in the practical applications which in turn affects the graphene based FETs and their reliability. Passivation layer is a key process to circumvent the above issues. Several passivation techniques like deposition of polymers, inorganic oxides and SAMs like Pentacene \cite{jee2009pentacene}, PECVD Si\textsubscript{3}N\textsubscript{4} \cite{lee2012multi}, Self Assembled Monolayer (SAM) \textit{viz.} 1-aminodecane \cite{long2012non}, H-BN \cite{mayorov2011micrometer} and ALD  Al\textsubscript{2}O\textsubscript{3} \cite{bzurutuzaaelorza2015highly, alexander2016encapsulation, kim2009realization,kang2013mechanism, vervuurt2017uniform, kim2018high} by few groups.

Among various deposition techniques, the sol-gel based deposition techniques have advantages due to its high deposition rate, no vacuum requirement, low cost method, tuning of the precursor composition \cite{brinker2013sol,cochran2019unique}. The quality of the sol-gel film will affect the transistor stability and overall sample yield. The two important parameters for any solution processed film are molarity of the film and annealing technique \cite{park2017sol}. The molarity of the film need to be optimized as higher or lower molarity of the film results in poor film quality in terms of density of the film \cite{khatibani2014growth} and shrinkage stress generated during densification of the film \cite{kozuka2003stress}.  The annealing technique and recipe plays an important role in densification process of the film \cite{liu2018solution}, as effective annealing technique leads to densification of the film at lower thermal budget with minimum shrinkage stress \cite{park2015depth}. Recently the sol-gel Alumina (Al\textsubscript{2}O\textsubscript{3}) has been extensively used in the solar cells as a field-effect passivation layer \cite{kalaivani2015spray} due to its good interface quality and dielectric breakdown. Park et.al \cite{park2016solution,kim2019direct} and Bae et.al \cite{bae2014fabrication} have explored sol-gel Alumina on the graphene transistors as dielectric layer and pH sensing membrane and none of them have explored for the passivation of the graphene transistors.

In our previous work, we have shown that DUV annealing technique is better than thermal annealing in terms of reduced thermal budget and lower shrinkage stress. We have also proposed two step seed layer deposition technique to resolve the crack issue in sol-gel Alumina film post annealing. In this work, we continue to use DUV annealing technique with two step seed layer deposition method. Here we are investigating process yield and device stability and propose a method to improve both using optimized molarity of the film. The sol-gel alumina film quality is studied on blanket wafer using UV-VIS, ellipsometry and XPS techniques. The detail electrical characterization (transfer characteristics, hysteresis and pulse hysteresis) and analysis (mobility and trap density extraction) of GFET devices is also presented in the work.

\section{Experimental }

\subsection{Synthesis of sol-gel Alumina precursor}

The sol-gel Alumina was prepared by dissolving the Aluminium nitrate nonahydrate (solute) (Al(NO\textsubscript{3})\textsubscript{3}.9H\textsubscript{2}O) (from Sigma Aldrich ) in Iso-propyl alcohol (solvent). The choice of the salt was due to its low thermal decomposition temperature (200\textsuperscript{o}C) \cite{cochran2019unique}. The resultant mixture was stirred at 1000 rpm at 55\textsuperscript{o}C for one hour to dissolve the salt completely in solvent. In addition to that, the chelating agent Acetyl-acetone was added to the above solution and stirred for 12 hours and the resultant solution was kept to aging for 3 days. The addition of chelating agent to the solution could reduce the residual stress during the annealing process \cite{kozuka2003stress}. These chelating coordination bonds hinder the hydrolysis and/or poly condensation, making the gel network flexible, and allowing structural relaxation in films during annealing and also enhance the DUV absorption \cite{kozuka2003stress}. The resultant solution was filtered using membrane filter 0.22 $\mu$m and used for the deposition on the fabricated devices. The solution was prepared for different molarities of 0.1 M and 0.05 M was spin coated on RCA cleaned silicon wafers followed by the drying at 110\textsuperscript{o}C for 15 minutes. The Deep Ultra violet (DUV) annealing (Figure \ref{schematic121} (c)) has been performed in UV Ozone cleaner (Jelight company Inc.) with intensity of 11 mW/cm\textsuperscript{2} in N\textsubscript{2} ambience for the densification of the sol-gel Alumina. The unintentional heating temperature during DUV annealing was around 70\textsuperscript{o}C and measured using the thermocouple inside the chamber during annealing process.

\subsection{Fabrication of graphene FETs}

\begin{figure*}[!ht]
\centering
\includegraphics[width= \textwidth]{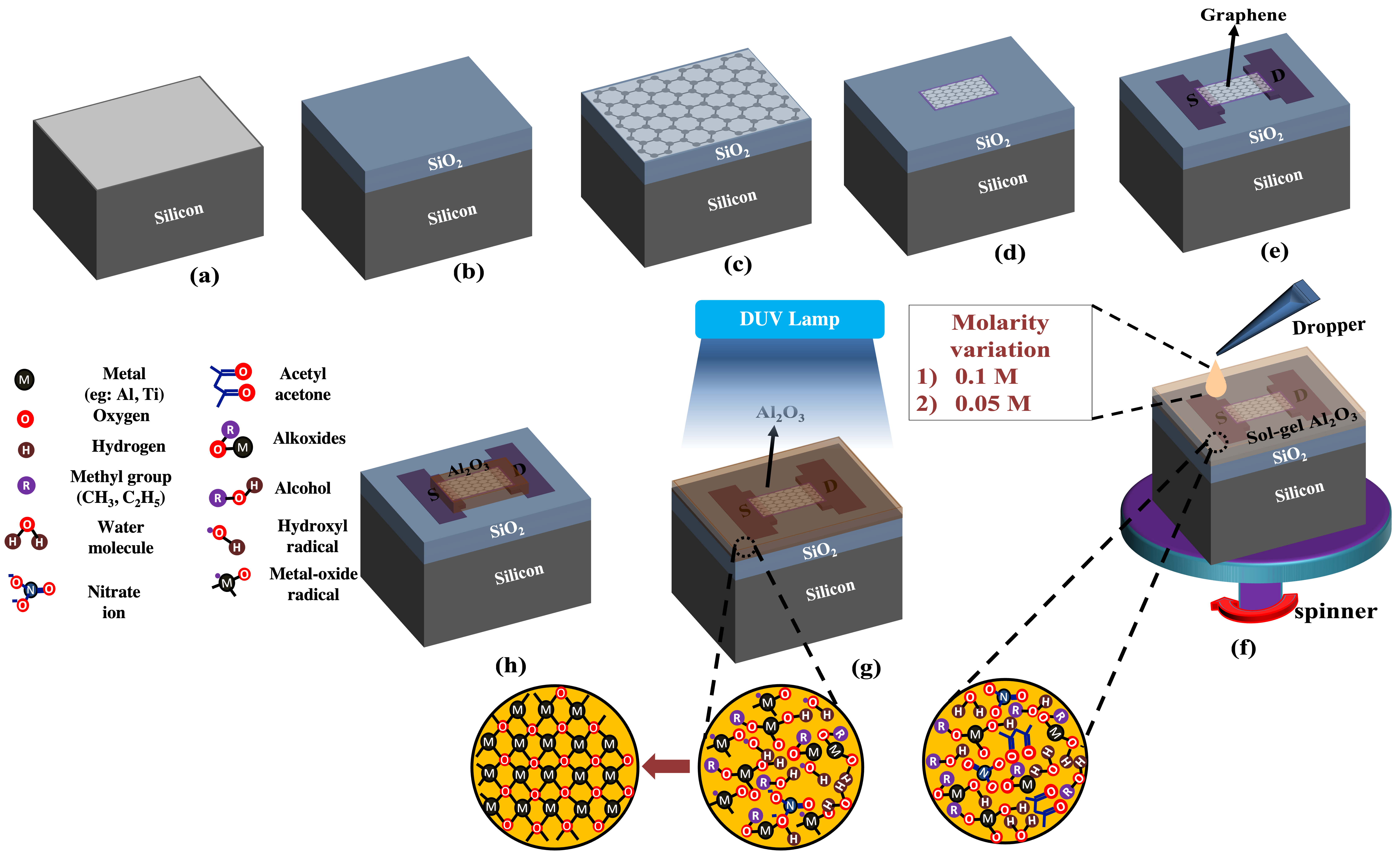}
\caption{\label{schematic121} The 3-D schematic process flow for sol-gel passivation layer based GFETs (a) RCA cleaned Silicon (b) Thermal oxidation of Silicon (c) CVD grown graphene transfer (d) Patterning of graphene (e) Contact deposition (f) Spin coating of sol-gel Alumina (g) DUV annealing of sol-gel Alumina (h) Patterning of sol-gel Alumina layer (the below circles shows the Zoom in version of the status of the sol-gel solution)}
\end{figure*}

The 3-D schematic process flow for sol-gel Alumina passivation layer based GFETs is shown in Figure \ref{schematic121}. We have purchased graphene monolayer grown on a copper foil using a CVD process from Graphenea Inc. Primarily, we prepared a carrier substrate to transfer the graphene layer from copper foil. An RCA cleaned 2-inch silicon wafer with resistivity 0.001-0.005 $\Omega$-cm, was used as a carrier substrate. To provide reliable isolation from the substrate and enable an optical contrast to spot the graphene monolayer, we have thermally grown 90 nm SiO\textsubscript{2} on the carrier wafer. Next, we have transferred the graphene monolayer on the SiO\textsubscript{2} layer. The large area Graphene monolayer was patterned to respective channel lengths and widths using photo-lithography followed by Oxygen plasma exposure for 5 min with RF power of 50 Watts. After the stripping of the resist (AZ 5214E) in AZ 100 remover, the sample was patterned for contacts. The stack of Ni/Au (20 nm / 30 nm) contact was deposited using an Electron beam evaporator at the 2.2X10\textsuperscript{-8} torr vacuum followed by lift-off. After that, the electrical characteristics of as-fabricated GFET were performed. Next, the sequential deposition of the seed layer of thickness 1.5 nm from the E-beam evaporation and 1.5 nm from the sputtering was performed. After that, the prepared sol-gel solution of different molarities was spin-coated at 5000 rpm for 45 seconds followed by drying at 110\textsuperscript{o}C. Then the DUV annealing was performed, followed by the patterning of the film for contacts opening using 7:1 BHF as shown in the Figure \ref{schematic121} (h).

\subsection{Characterization}
The Microscope images were taken using an Olympus microscope. The Scannig Electron Microscopy (SEM) image was performed using JEOL JSM-7600F FEG-SEM. The Atomic Force Microscopy (AFM) was performed using tapping mode using MFP-3D AFM, Asylum research, Oxford instruments. The UV-VIS spectroscopy was performed on the sol-gel Alumina solution using a Perkin-Elmer UV-VIS-NIR spectrometer. The X-ray Photoelectron Spectroscopy (XPS) of the sol-gel Alumina film samples was studied using PHI 5000 versa probe II with monochromatic Al K$\alpha$ X-ray source. The thickness measurement was analyzed using an ellipsometer from SENTECH instruments. The electrical characterization on the respective samples was done using a Keysight B1500A semiconductor device analyzer.

\section{Results and discussion}

\subsection{Process yield and device reliability study at 0.1 M}

\begin{figure}[!ht]
\centering
\includegraphics[width= \columnwidth]{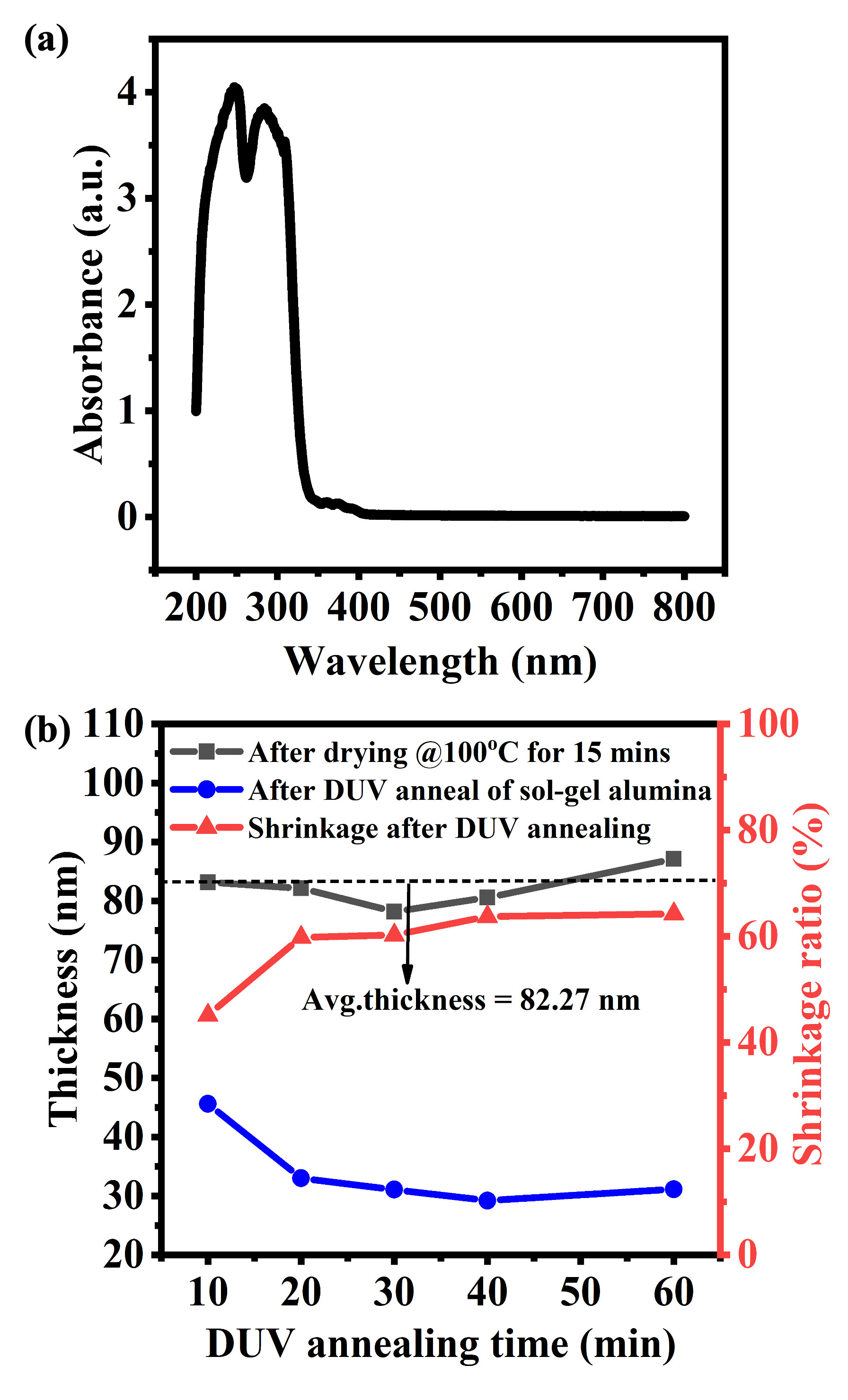}
\caption{\label{0.1M-uvvis} (a) UV-VIS spectroscopy of sol-gel Alumina (b) Ellipsometry based thickness measurement and shrinkage ratio extraction for different annealing times }
\end{figure}

To check if DUV is an effective annealing technique for sol-gel Alumina film, the absorbance of precursor solution was carried out using UV-VIS spectroscopy between 200-800 nm (Figure \ref{0.1M-uvvis} (a)). The absorbance peak in the range of 200-300 nm confirms that DUV will be an effective technique to anneal sol-gel Alumina film. To further find out the optimized DUV annealing time which can completely remove solvent from the film, annealing time splits were carried out on blanket deposited film. The pre and post-annealing film thicknesses were measured using ellipsometry, and the shrinkage ratio was extracted as shown in Figure \ref{0.1M-uvvis} (b). The thickness reduction of sol-gel is due to the accelerated polycondensation process due to the formation of OH radicals \cite{van1995uv,park2015depth}. In the reported work, the Aluminium nitrate nonahydrate salt will dissociate and give NO\textsubscript{2} and OH radicals as shown below.

It can be seen from Figure \ref{0.1M-uvvis} (b) that (i) average thickness of pre-annealing films across different samples is 82 nm, (ii) as the annealing time increases, the film thickness reduces till 40 minutes, beyond which the film thickness got saturated to ~30 nm, and there was no more thickness reduction and (iii) the shrinkage ratio of 0.1 M film is close to 60$\%$, which means that there was a significant amount of solvent present in the film and hence it might be taking longer time to remove the solvent altogether. This could lead to higher shrinkage stress due to a longer annealing time.

\begin{figure}[!ht]
\centering
\includegraphics[width= \columnwidth]{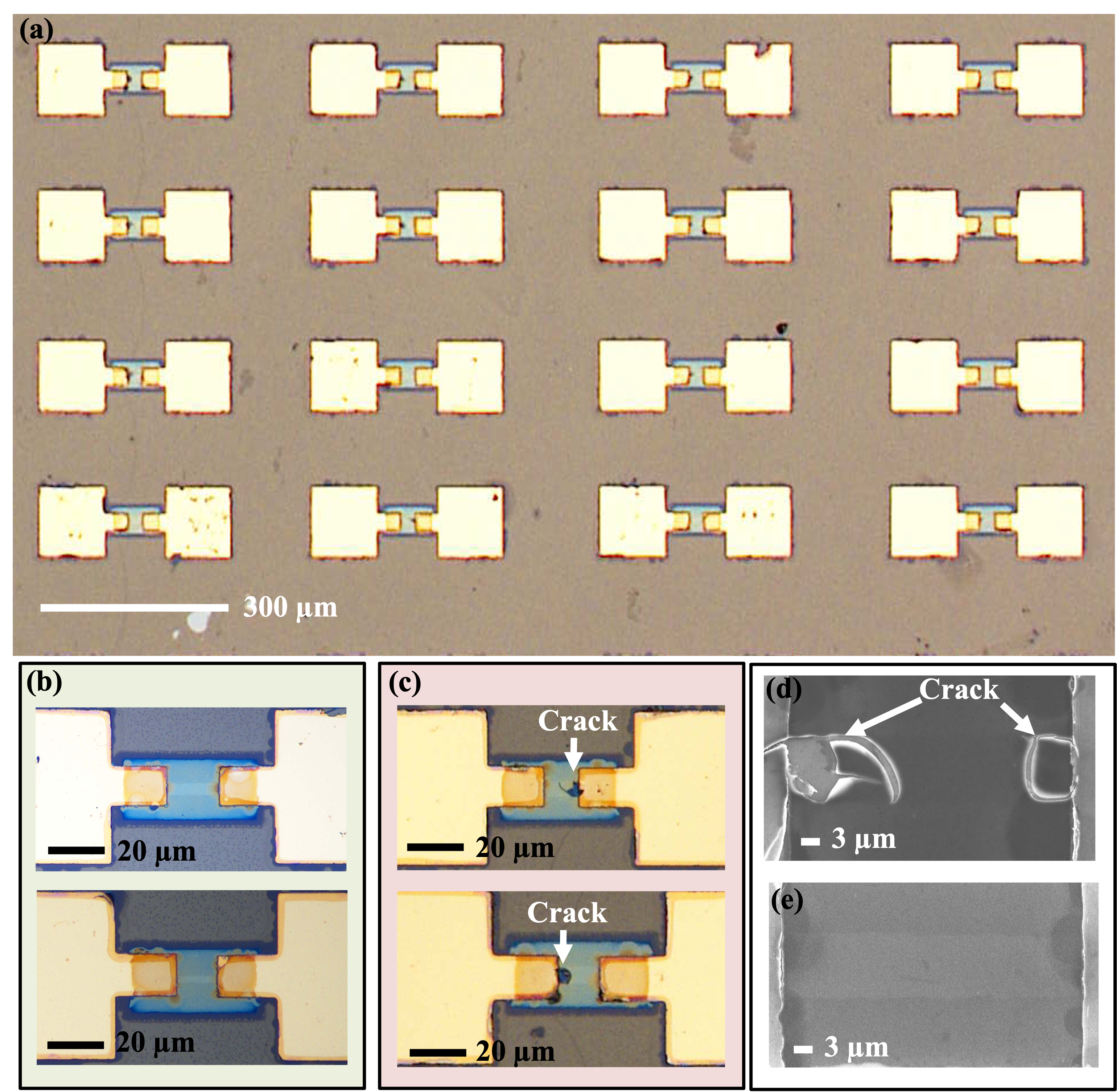}
\caption{\label{0.1Mfig} Optical microscopic picture of (a) GFETs array on sample (b) zoomed image showing devices without any crack in channel region and (c) zoomed image of devices showing crack in the channel region. (d) and (e) are SEM images of non-crack (above) and crack (below) devices respectively}
\end{figure}

The GFETs were fabricated with DUV annealing of 20 minutes. The optical microscopic picture of sample with array of GFETs was shown in Figure \ref{0.1Mfig} (a). 
When the devices were inspected at 50X, it was found that some of the devices had crack (Figure \ref{0.1Mfig} (e)) in the sol-gel film post-annealing, and some of them did not have the crack (Figure \ref{0.1Mfig} (d)). The devices were also inspected, and crack formation was confirmed using SEM images shown in Figure \ref{0.1Mfig} (e) (without crack) and Figure \ref{0.1Mfig} (f) (with crack). This crack formation could be due to the combined effect of inadequate densification of the sol-gel precursor, insufficient removal of the solvents and the stress due to shrinkage during DUV annealing.

\begin{figure}[!ht]
\centering
\includegraphics[width= \columnwidth]{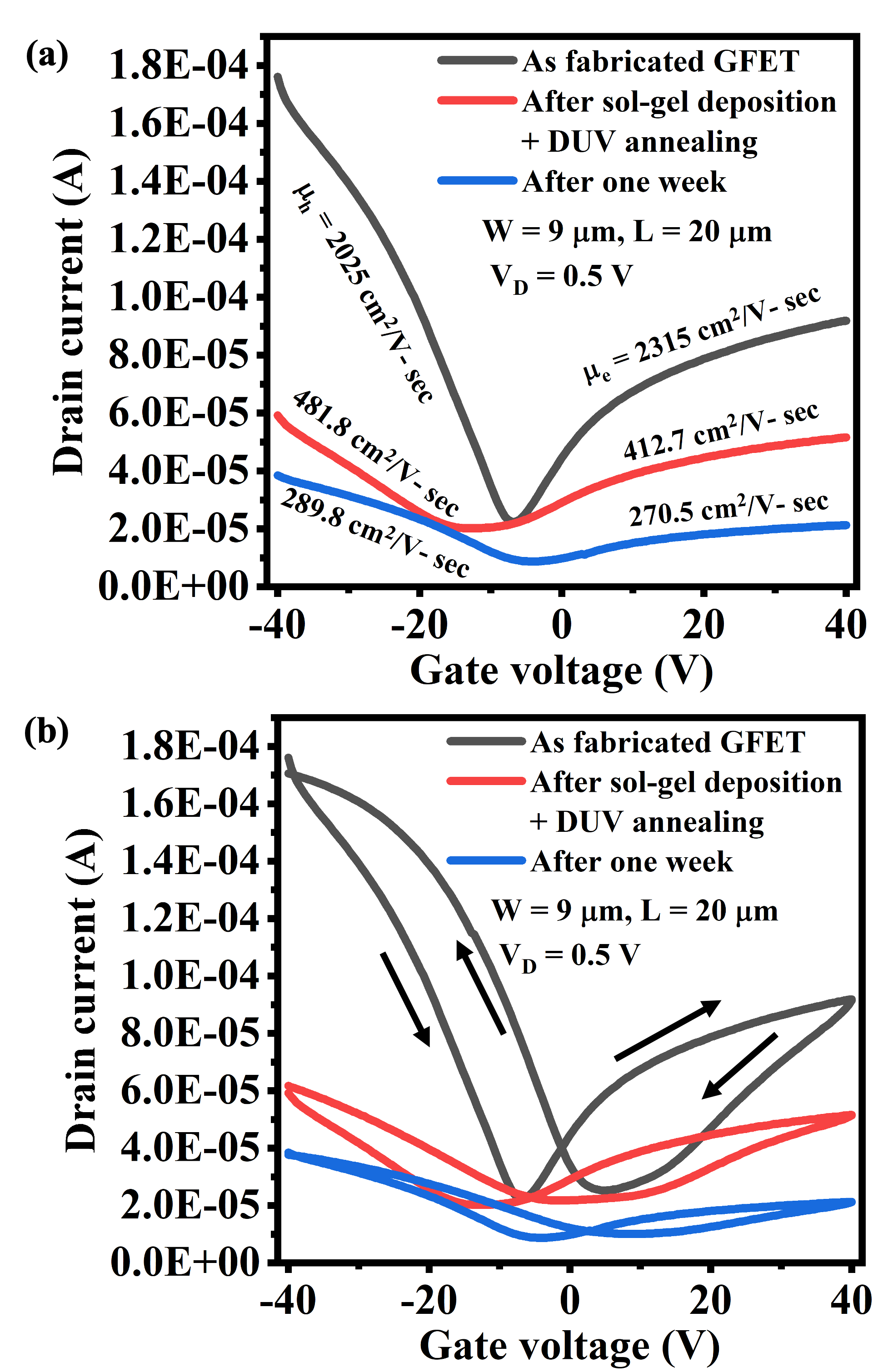}
\caption{\label{0.1IV fig} (a) DC-IV characteristics of GFET before and after sol-gel deposition with 0.1 M molarity (b) DC-IV hysteresis characteristics before and after Alumina deposition with 0.1 M molarity}
\end{figure}

Figure \ref{0.1IV fig} (a) shows the DC transfer characteristics of the graphene FET before and after sol-gel deposition. The mobilities of both hole and electron branches were found to be 2025 cm\textsuperscript{2}/V-sec and 2315 cm\textsuperscript{2}/V-sec for as-fabricated GFET. The mobilities have been extracted from the FTM method \cite{zhong2015comparison}. The As fabricated GFET shows n-type doping which was shown as a shift in the Dirac point towards negative voltage i.e. at -6.8 V, which could be due to photo-resist stripper \cite{sul2016reduction} and the asymmetry in transfer characteristics and lower conductance in the electron branch could be due to work-function difference between contacts and graphene \cite{nagashio2009metal}.

After sol-gel Alumina deposition followed by DUV annealing, the hole and electron mobilities were degraded and have been found to be 481.8 cm\textsuperscript{2}/V-sec and 412.7 cm\textsuperscript{2}/V-sec. This could be due to increase of the charge impurity scattering \cite{chen2008charged}. These scatters were formed due to the OH ion generation during the redox reaction from the diffusion of O\textsubscript{2} and H\textsubscript{2}O from the atmosphere through the grain boundaries of sol-gel Alumina layer, or it could be due to trapped solvents in the metal oxide network which could act as a remote coulomb scatters. 
The trapped solvents could also form the redox reaction with the electrons at the graphene interface leading to OH scatters causing the degradation \cite{kang2013mechanism}. The transfer characteristics got further degraded after one week, causing the degradation of hole and electron branch with mobilities of 289.8 cm\textsuperscript{2}/V-sec and 270.5 cm\textsuperscript{2}/V-sec. This could be due to diffusion of the H\textsubscript{2}O and O\textsubscript{2} from the ambient atmosphere through the grain boundaries \cite{kang2013mechanism} of sol-gel Alumina layer, causing more degradation with time which can also be supported by a shift in Dirac point towards positive voltage direction showing P-type nature. The width of DC-IV characteristics around Dirac point ($\Delta$V\textsubscript{g}\textsubscript{-min}) broadens after one week which could be due to increase in concentration of charge impurities (n\textsubscript{imp}) \cite{chen2008charged} and the values of $\Delta$V\textsubscript{g}\textsubscript{-min} were extracted by intersecting a line through the minimum conductivity around Dirac point \cite{katoch2012impact} as shown in the table \ref{01 M extract}. 

The minimum conductivity ($\sigma$\textsubscript{min}) reduction after DUV annealing and one week also reveals that the sample has becoming dirtier as per adam et.al \cite{adam2007self} which means there were more concentration of charged impurities (n\textsubscript{imp}) \cite{adam2007self}

To further explore, the hysteresis has been explored as shown in Figure \ref{0.1IV fig} (b). The hysteresis characteristics show the positive direction of hysteresis \cite{wang2010hysteresis} before and after DUV annealed sol-gel Alumina layer. The hysteresis could be due to both redox and charge tunneling components \cite{lee2013quantitative}. The hysteresis has increased after DUV annealing treatment which can be anticipated due to an increase in both the redox and tunneling components \cite{lee2013quantitative}. This has been shown through the charge trap density at the graphene interface from the equation N\textsubscript{trap}=C\textsubscript{ox}*($\Delta$V\textsubscript{Dirac})/q where C\textsubscript{ox} os the SiO\textsubscript{2} oxide capacitance and $\Delta$V\textsubscript{Dirac} is the difference in the voltages at the Dirac point. The charge trap density for the as fabricated GFET was 2.9 $\times$ 10\textsuperscript{12} /cm\textsuperscript{2} for the $\Delta$V\textsubscript{Dirac} of 12.4 V and after the DUV annealing was 4.5 $\times$ 10\textsuperscript{12} /cm\textsuperscript{2} for the  $\Delta$V\textsubscript{Dirac} of 19.2 V. However, after one week, the hysteresis has reduced i.e. $\Delta$V\textsubscript{Dirac} of 14.4 V resulting charge trap density of 3.37 $\times$ 10\textsuperscript{12} /cm\textsuperscript{2}, which could be due to partial reduction of tunneling component i.e. the charge trapping (fast component) into adsorbates (H\textsubscript{2}O) \cite{lee2013quantitative} which means the H\textsubscript{2}O has been utilized for the formation of OH\textsuperscript{--} ion generation through redox reaction (slow component), which has degraded the mobility as shown in the Figure \ref{0.1IV fig} (a). The extracted parameters from transfer and hysteresis characteristics have been tabulated in table \ref{01 M extract}.

\begin{table}
\centering
\caption{Parameter extraction of GFET with 0.1 M sol-gel Alumina ($\mu$\textsubscript{h}, $\mu$\textsubscript{e}, $\Delta$V\textsubscript{g}\textsubscript{-min}, $\sigma$\textsubscript{min}, $\Delta$V\textsubscript{Dirac} and N\textsubscript{trap} denotes the Hole mobility, Electron mobility, Minimum conductivity plateau width around Dirac point, minimum conductivity at Dirac point, Dirac point difference during dual sweep and charge trap density)}

\begin{tabular}{|c|ccc|}
\hline
                                                                                    & \multicolumn{3}{c|}{\textbf{Molarity - 0.1 M}}                                                                                                                         \\ \cline{2-4} 
 {\textbf{Parameters }}                                            & \multicolumn{1}{c|}{\textbf{\begin{tabular}[c]{@{}c@{}}As \\ fabricated\\ GFET\end{tabular}}} & \multicolumn{1}{c|}{\textbf{\begin{tabular}[c]{@{}c@{}}After \\ DUV \\ annealing\end{tabular}}} & \textbf{\begin{tabular}[c]{@{}c@{}}After one \\ week\end{tabular}} \\ \hline
\textbf{\begin{tabular}[c]{@{}c@{}}$\mu$\textsubscript{h}\\ (cm\textsuperscript{2}/V-sec)\end{tabular}}     & \multicolumn{1}{c|}{\cellcolor[HTML]{FFFFFF}{\color[HTML]{000000} 2025}}& \multicolumn{1}{c|}{\cellcolor[HTML]{FFFFFF}{\color[HTML]{000000} 481.5}}                       & \cellcolor[HTML]{FFFFFF}{\color[HTML]{000000} 289.5}               \\ \hline
\textbf{\begin{tabular}[c]{@{}c@{}}$\mu$\textsubscript{e} \\ (cm\textsuperscript{2}/V-sec)\end{tabular}} &

\multicolumn{1}{c|}{\cellcolor[HTML]{FFFFFF}{\color[HTML]{000000} 2315}}& \multicolumn{1}{c|}{\cellcolor[HTML]{FFFFFF}{\color[HTML]{000000} 412.7}}                       & \cellcolor[HTML]{FFFFFF}{\color[HTML]{000000} 270.5}               

\\ \hline

\textbf{\begin{tabular}[c]{@{}c@{}}$\Delta$V\textsubscript{g}\textsubscript{-min} (V)\end{tabular}}     & \multicolumn{1}{c|}{5.4}                                                                      & \multicolumn{1}{c|}{8.4}                                                                        & 12.3                                                               \\ \hline
\textbf{\begin{tabular}[c]{@{}c@{}}$\sigma$\textsubscript{min}\end{tabular}}             & \multicolumn{1}{c|}{\cellcolor[HTML]{FFFFFF}{\color[HTML]{000000} 4.75 q\textsuperscript{2}/h}}                 & \multicolumn{1}{c|}{\cellcolor[HTML]{FFFFFF}{\color[HTML]{000000} 1.45 q\textsuperscript{2}/h}}                   & \cellcolor[HTML]{FFFFFF}{\color[HTML]{000000} 1.37 q\textsuperscript{2}/h}           \\ \hline

\textbf{\begin{tabular}[c]{@{}c@{}}$\Delta$V\textsubscript{Dirac} (V)\end{tabular}}                     & \multicolumn{1}{c|}{\cellcolor[HTML]{FFFFFF}{\color[HTML]{000000} 12.4}}& \multicolumn{1}{c|}{\cellcolor[HTML]{FFFFFF}{\color[HTML]{000000} 19.2}}                        & \cellcolor[HTML]{FFFFFF}{\color[HTML]{000000} 14.4}                \\ \hline
\textbf{\begin{tabular}[c]{@{}c@{}}N\textsubscript{trap} (/cm\textsuperscript{2})\end{tabular}} & \multicolumn{1}{c|}{\cellcolor[HTML]{FFFFFF}{\color[HTML]{000000} 2.9e12}}& \multicolumn{1}{c|}{\cellcolor[HTML]{FFFFFF}{\color[HTML]{000000} 4.5e12}}                      & \cellcolor[HTML]{FFFFFF}{\color[HTML]{000000} 3.37e12}             \\ \hline

\end{tabular}

\label{01 M extract}
\end{table}

Based on the above discussion, there were two problems, (i) process yield was not good as many devices shows crack in sol-gel Alumina layer above graphene channel and (ii) device stability is not good, and keeps degrading in terms of absolute current and mobility. In order to resolve the issues, we propose sol-gel Alumina with low molarity (0.05M)

\subsection{Process yield and device reliability study at 0.05 M}
\begin{figure}[!ht]
\centering
\includegraphics[width= \columnwidth]{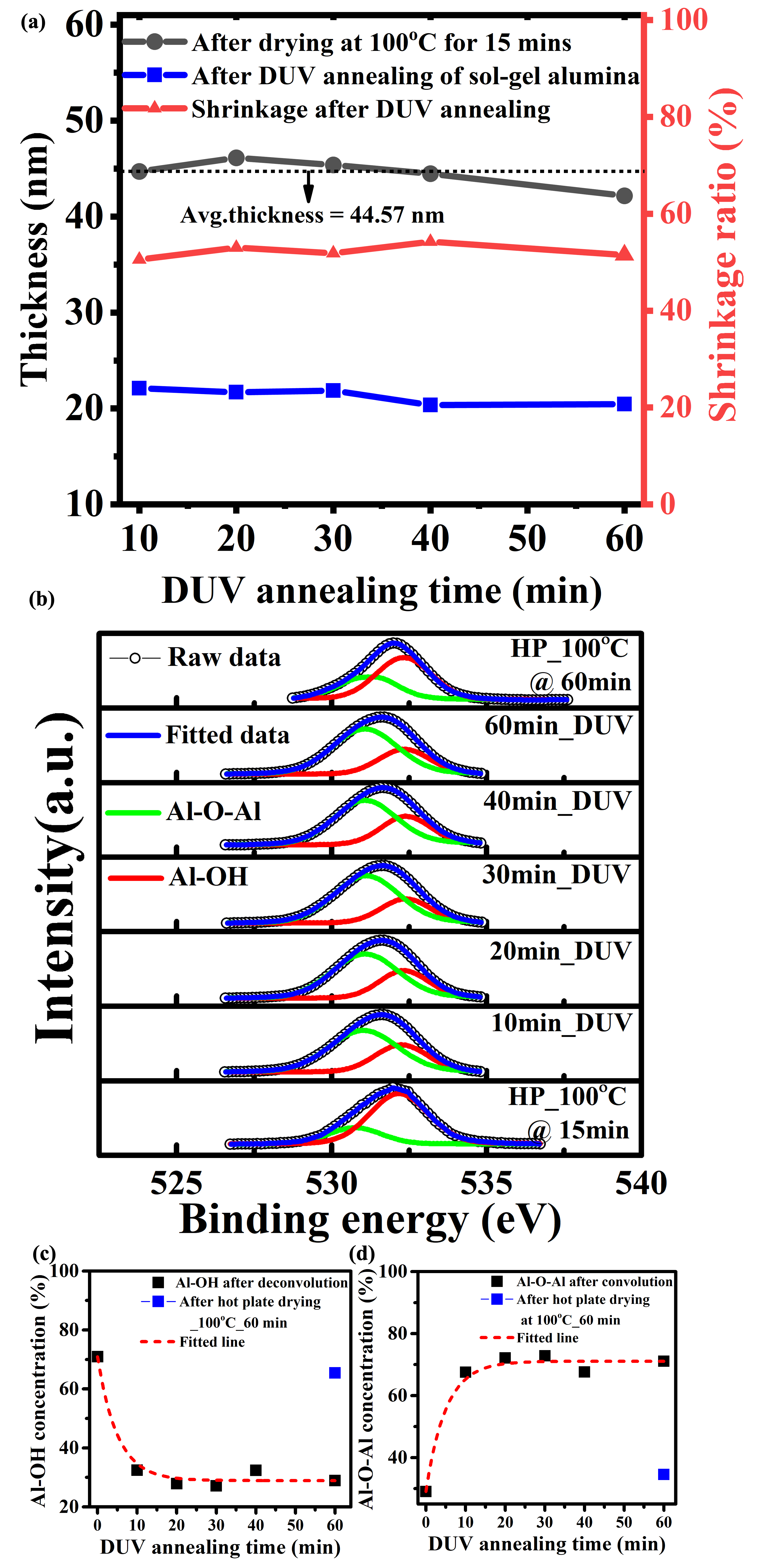}
\caption{\label{xps1}  (a) Annealing time optimization and shrinkage ratio after DUV Annealing for different times for 0.05 M (b) XPS: Oxygen deconvolution for different DUV annealing time (For reference, the hot plate drying is carried at 100\textsuperscript{o}C for 15 minutes and also 60 minutes) (c) Al-O-Al concentration Vs. the DUV annealing time (d) Al-OH concentration Vs. the DUV annealing time}
\end{figure}
To circumvent the shrinkage stress with 0.1 M, the sol-gel Alumina molarity has been reduced to 0.05 M. Figure \ref{xps1} (a) shows the annealing time optimization of the 0.05 M sol-gel Alumina respectively with different DUV annealing time with the help of ellipsometry. 
The average thickness for 0.05 M solution was ~44.57 nm measured after the drying at 100\textsuperscript{o}C. The final thickness obtained after DUV annealing time was around ~22 nm. The shrinkage ratio with DUV annealing time was also plotted, showing that the thickness got saturated at 50$\%$ at 10 minutes showing less shrinkage stress during DUV annealing.
To explore the stochiometry of sol-gel Alumina after the DUV annealing, XPS measurement was performed on the samples and oxygen (O 1s) peak is deconvoluted for Al-O-Al peak and Al-OH at 531.0 and 532.3 respectively, as shown in the Figure \ref{xps1} (b). After the deconvolution, the Al-O-Al concentration and Al-OH concentration is plotted as a function of DUV annealing time as shown in Figure \ref{xps1} (c) and (d). The zero minute sample in Figure \ref{xps1} (c) and (d) is as-spun coated followed by drying on a hotplate at 100\textsuperscript{o}C for 15 minutes. 
From the graph \ref{xps1} (c) and (d), it was conveyed that the Al-O-Al and Al-OH concentration is almost constant after the 20 minutes. To investigate the effect and advantage of DUV annealing, the sample was kept on a hot plate at 100\textsuperscript{o}C for 60 minutes.

\begin{figure}[!ht]
\centering
\includegraphics[width= \columnwidth]{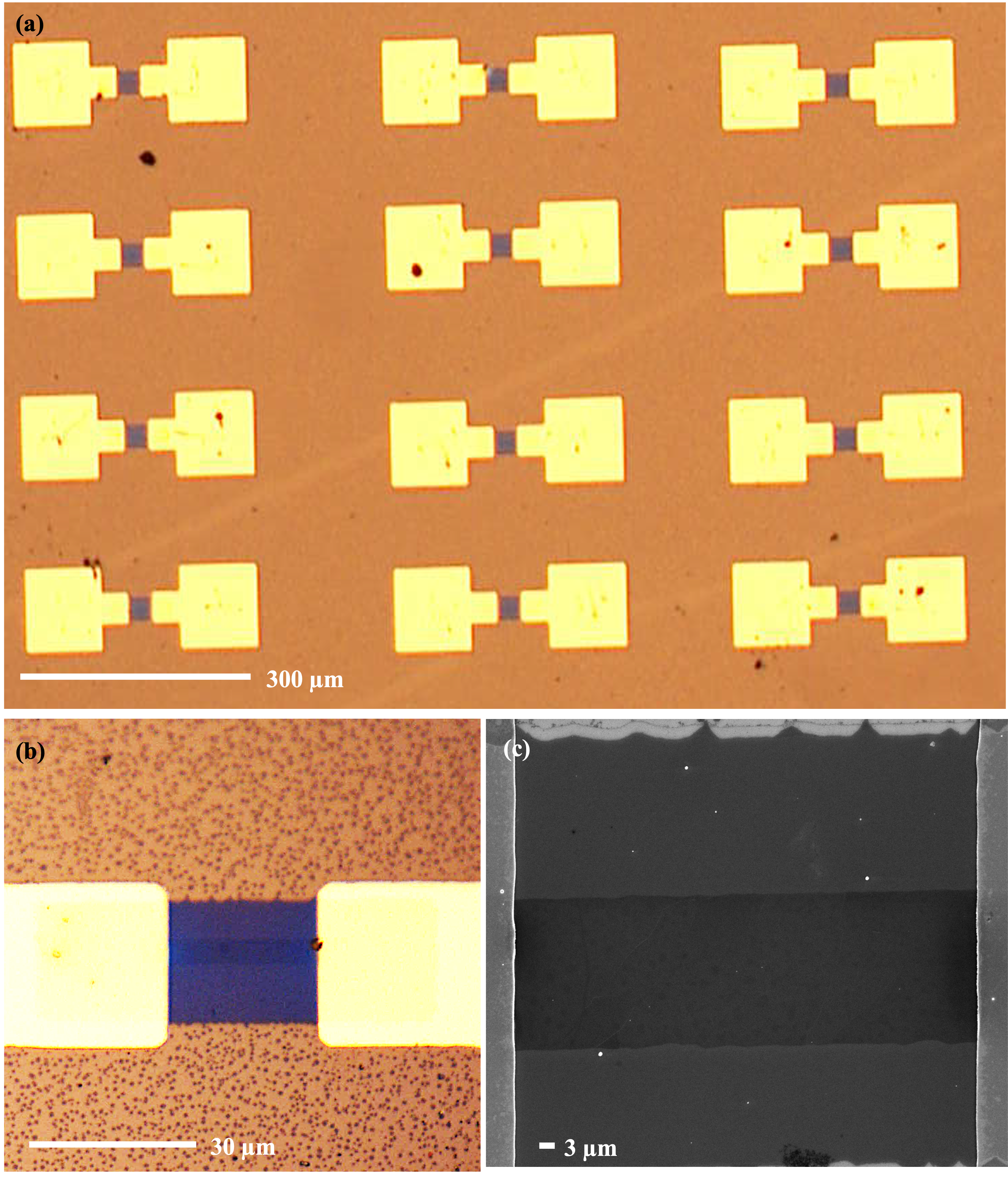}
\caption{\label{0.05Mfig} (a) Optical microscope image GFETs devices array with sol-gel Alumina as a passivation (b) Zoomed optical microscope image after sol-gel deposition followed by DUV annealing for 20 minutes (c) SEM images of passivated GFET }
\end{figure}

\begin{figure*}[!ht]
\centering
\includegraphics[width= \textwidth]{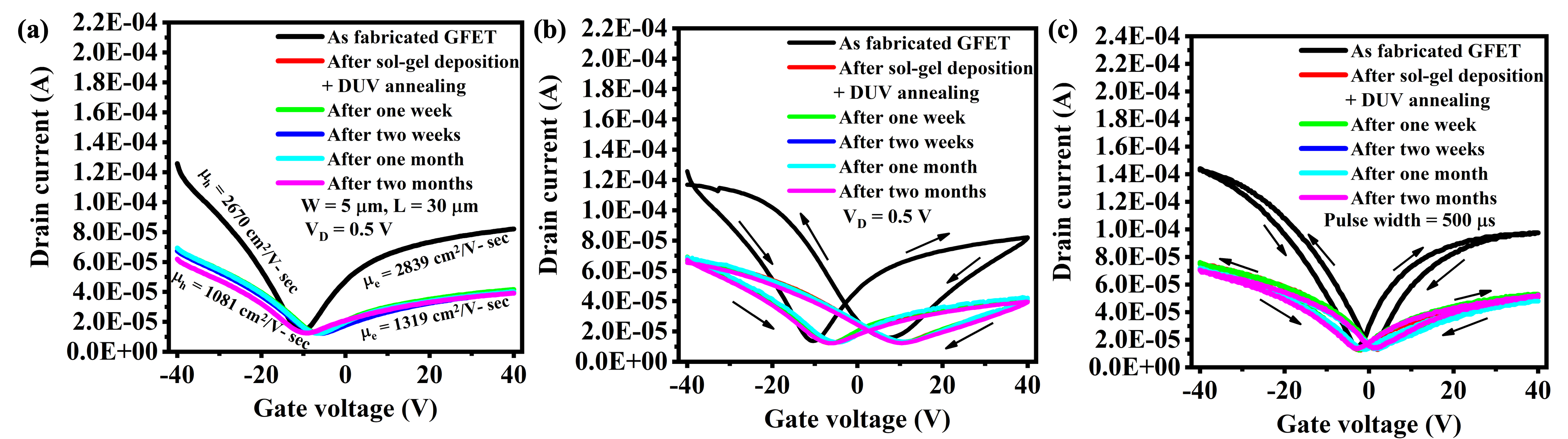}
\caption{\label{0.05IVfig} (a) DC-IV characteristics of GFET before and after sol-gel deposition with 0.05 M molarity (b) DC-IV hysteresis characteristics before and after Alumina deposition with 0.05 M molarity (c) Pulse-IV hysteresis characteristics before and after Alumina deposition with pulse width of 500 $\mu$s}
\end{figure*}

During the DUV annealing process, the sample will undergo unintentional heating to around 75-80 \textsuperscript{o}C. Still, to see the effect of DUV annealing on the densification process, the hot plate treated sample for 60 min at 100\textsuperscript{o}C was investigated to have a fair comparison. It didn't show much change in the Al-O-Al and Al-OH concentration even after 60 minutes. However, the DUV annealed sample at 60 minutes showed the reduction in Al-OH concentration as shown in Figure \ref{xps1} (c) and (d). 

Optical image in Figure \ref{0.05Mfig} (a) shows sample with 100 $\%$ yield concerning crack removal in the sol-gel Alumina layer on the graphene channel. This could be due to the reduction of shrinkage stress component with 0.05 M. Figure \ref{0.05Mfig} (b) shows the enlarged version of one of the devices showing no crack in the active region. This can be further confirmed from the SEM, as shown in the Figure \ref{0.05Mfig} (c).

Figure \ref{0.05IVfig} (a) shows the DC transfer characteristics of GFET before and after sol-gel deposition. The As fabricated GFET shows n-type doping, which has been established as a shift in the Dirac point towards negative voltage, i.e., at -10.8 V, which could be due to resist stripper \cite{sul2016reduction} and the asymmetry in transfer characteristics and lower conductance in the electron branch was due to the work-function difference \cite{nagashio2009metal}. 

After the DUV annealing of sol-gel Alumina for 20 minutes, there is a slight shift in the Dirac point. The decrease of the mobility in electron and hole branch, which inherently decreases current, the broadening of minimum conductivity plateau in the transfer characteristics have been observed. The shift in the Dirac point was at -6.4 eV, which shows the graphene FET is still n-type doped. The mobility has been extracted from the FTM method \cite{zhong2015comparison}. The hole and electron mobility before sol-gel deposition were 2670 cm\textsuperscript{2}/V-sec and 2839 cm\textsuperscript{2}/V-sec. The hole and electron mobility after the sol-gel deposition were 1081 cm\textsuperscript{2}/V-sec and 1319 cm\textsuperscript{2}/V-sec, and the same mobilities were maintained for two months. This shows that the 0.05 M sol-gel Alumina layer does not degrade the hole and electron mobilities compared to the 0.1 M sol-gel Alumina layer. Hence the device stability has improved compare to 0.1 M based GFETs

Figure \ref{0.05IVfig} (b) shows the hysteresis transfer characteristics that show the positive direction of hysteresis \cite{wang2010hysteresis}.
The hysteresis has been contributed from charge tunneling into the adsorbates (fast component) and redox couple reaction (slow component) \cite{lee2013quantitative}.
The charge trap density for the as fabricated GFET was 4.21 $\times$ 10\textsuperscript{12} /cm\textsuperscript{2} for the $\Delta$V\textsubscript{Dirac} of 18 V and after the  DUV annealing was 4.02 $\times$ 10\textsuperscript{12} /cm\textsuperscript{2} for the  $\Delta$V\textsubscript{Dirac} of 17.2 V. The minute reduction in charge trap density after the DUV annealing could be due to removal of water and oxygen to some extent from the oxide network.

\begin{table}
\centering
\caption{Parameter extraction of GFET with 0.05 M sol-gel Alumina ($\mu$\textsubscript{h}, $\mu$\textsubscript{e}, $\Delta$V\textsubscript{g}\textsubscript{-min}, $\sigma$\textsubscript{min}, $\Delta$V\textsubscript{Dirac} and N\textsubscript{trap} denotes the Hole mobility, Electron mobility, Minimum conductivity plateau width around Dirac point, minimum conductivity at Dirac point, Dirac point difference during dual sweep and charge trap density)}
\label{005extract}
\begin{tabular}{|c|c|c|c|} 
\hline
{\textbf{Parameters}}                                                                    & \multicolumn{3}{c|}{\textbf{Molarity - 0.05 M}}                                                                                                                                                                                                                                                                                  \\ 
\cline{2-4}
                                                                                                        & \begin{tabular}[c]{@{}c@{}}\textbf{As }\\\textbf{ fabricated}\\\textbf{ GFET}\end{tabular} & \begin{tabular}[c]{@{}c@{}}\textbf{After }\\\textbf{ DUV }\\\textbf{ anneal}\end{tabular} & \begin{tabular}[c]{@{}c@{}}\textbf{After one}\\\textbf{ week/two }\\\textbf{ weeks/one/}\\\textbf{ two}\\\textbf{ months}\end{tabular}  \\ 
\hline
\begin{tabular}[c]{@{}c@{}}\textbf{$\mu$\textsubscript{h}}\\\textbf{ (cm\textsuperscript{2}/V-sec)}\end{tabular}        & 2670                                                                                       & 1081                                                                                      & 1081                                                                                                                                    \\ 
\hline
\begin{tabular}[c]{@{}c@{}}\textbf{$\mu$\textsubscript{e} }\\\textbf{ (cm\textsuperscript{2}/V-sec)}\end{tabular}       & 2839                                                                                       & 1319                                                                                      & 1319                                                                                                                                    \\ 
\hline
\textbf{$\Delta$V\textsubscript{g-min} (V)}                                                                             & 3.2                                                                                        & 5.9                                                                                       & 5.9                                                                                                                                     \\ 
\hline
\textbf{$\sigma$\textsubscript{min}}                                                                                    & 4.4 q\textsuperscript{2}/h                                                                 & 4.0 q\textsuperscript{2}/h                                                                & 4.0 q\textsuperscript{2}/h                                                                                                              \\ 
\hline
\begin{tabular}[c]{@{}c@{}}\textbf{DC-IV}\\\textbf{ $\Delta$V\textsubscript{Dirac} (V)}\end{tabular}                    & 18                                                                                         & 17.2                                                                                      & 17.2                                                                                                                                    \\ 
\hline
\begin{tabular}[c]{@{}c@{}}\textbf{N\textsubscript{trap} (/cm\textsuperscript{2})}\\\textbf{ (DC-IV)}\end{tabular}      & 2.9e12~                                                                                    & 4.5e12~                                                                                   & 3.37e12                                                                                                                                 \\ 
\hline
\begin{tabular}[c]{@{}c@{}}\textbf{Pulse - IV}\\\textbf{ $\Delta$V\textsubscript{Dirac} (V)}\end{tabular}               & 3.6                                                                                        & 4.4                                                                                       & 4.4                                                                                                                                     \\ 
\hline
\begin{tabular}[c]{@{}c@{}}\textbf{N\textsubscript{trap} (/cm\textsuperscript{2})}\\\textbf{ (Pulse - IV)}\end{tabular} & 8.42e11                                                                                    & 1.02e12                                                                                   & 1.02e12                                                                                                                                 \\
\hline
\end{tabular}
\end{table}

In order to further explore about the hysteresis, pulse IV characterisation has been performed with SMU pulse with the pulse width of 500 $\mu$sec on graphene FET before and after sol-gel Alumina deposition as shown in the Figure \ref{0.05IVfig} (c). The charge trap density for the as fabricated GFET was 8.42 $\times$ 10\textsuperscript{11} /cm\textsuperscript{2} for the $\Delta$V\textsubscript{Dirac} of 3.6 V and after the Alumina deposition charge trap density was 1.02 $\times$ 10\textsuperscript{12} /cm\textsuperscript{2} for the $\Delta$V\textsubscript{Dirac} of 4.4 V which shows less hysteresis. As the pulse IV characterization is time dependant measurement and the pulse width of 500 $\mu$s. It will bypass the slow component (redox component), leaving the fast component leading to more negligible hysteresis. The extracted parameters from transfer and hysteresis characteristics have been tabulated in table \ref{005extract}

\begin{table*}
\centering
\caption{Benchmark table for different Alumina deposition on GFETs and their extracted mobilities}
\label{tab:benchmarktable}
\begin{tabular}{|c|c|c|c|c|c|cc|cc|}
\hline
 {\textbf{Ref}} & {\textbf{\begin{tabular}[c]{@{}c@{}}Graphene \\ type\end{tabular}}} & {\textbf{\begin{tabular}[c]{@{}c@{}}Type of \\ gate \end{tabular}}} & {\textbf{\begin{tabular}[c]{@{}c@{}}W/L\\ ($\mu$m/$\mu$m)\end{tabular}}} & {\textbf{\begin{tabular}[c]{@{}c@{}}Type of\\deposition\end{tabular}}}    & {\textbf{\begin{tabular}[c]{@{}c@{}} Process \\temperature\\ (\textsuperscript{o}C)\end{tabular}}} & \multicolumn{2}{c|}{\textbf{\begin{tabular}[c]{@{}c@{}}Mobility\\ (cm\textsuperscript{2}/V-sec)\\ (Before\\ deposition)\end{tabular}}} & \multicolumn{2}{c|}{\textbf{\begin{tabular}[c]{@{}c@{}}Mobility \\ (cm\textsuperscript{2}/V-sec)\\  (After \\ deposition) \end{tabular}}} \\ \cline{7-10} 
                                                                &                                                                                    &                                                                                         &                                                                                       &                                                                                           &                                                                                     & \multicolumn{1}{c|}{\textbf{$\mu$\textsubscript{h}}}                                     & \textbf{$\mu$\textsubscript{e}}                                    & \multicolumn{1}{c|}{\textbf{$\mu$\textsubscript{h}}}                                                      & \textbf{$\mu$\textsubscript{e}}                                                      \\ \hline
\cite{bae2014fabrication}                                       & CVD                                                                                & \begin{tabular}[c]{@{}c@{}}Back\\ gate\end{tabular}                            & 10/10                                                                                 & \begin{tabular}[c]{@{}c@{}}Sol-gel \\ Alumina\\ +\\ Hot plate \\ annealing\end{tabular}   & 250                                                                                 & \multicolumn{1}{c|}{NA}                                              & NA                                             & \multicolumn{1}{c|}{550}                                                              & NA                                                               \\ \hline
\cite{park2016solution}                                         & \begin{tabular}[c]{@{}c@{}}Mechanical\\ exfoliation\end{tabular}                   & \begin{tabular}[c]{@{}c@{}}Top\\ gate\end{tabular}                     & 40/2                                                                                  & \begin{tabular}[c]{@{}c@{}}Sol-gel \\ Alumina\\ +\\ Hot plate \\ annealing\end{tabular}   & 250                                                                                 & \multicolumn{1}{c|}{NA}                                              & NA                                             & \multicolumn{1}{c|}{NA}                                                               & \begin{tabular}[c]{@{}c@{}}9000\\ \end{tabular} \\ \hline
\cite{kim2019direct}                                            & \begin{tabular}[c]{@{}c@{}}Epitaxial\\ growth on \\ SiC\end{tabular}               & \begin{tabular}[c]{@{}c@{}}Top\\ gate\end{tabular}                     & 1/1                                                                                   & \begin{tabular}[c]{@{}c@{}}Sol-gel \\ Alumina\\ +\\ Microwave \\ annealing\end{tabular}   & 500                                                                                 & \multicolumn{1}{c|}{NA}                                              & NA                                             & \multicolumn{1}{c|}{\begin{tabular}[c]{@{}c@{}}8.6 \\ \end{tabular}} & \begin{tabular}[c]{@{}c@{}}9.7\\  \end{tabular}   \\ \hline
\cite{bzurutuzaaelorza2015highly}                               & CVD                                                                                & \begin{tabular}[c]{@{}c@{}}Back\\ gate\end{tabular}                            & 10/25                                                                                 & \begin{tabular}[c]{@{}c@{}}Thermal\\ ALD\end{tabular}                                     & 150                                                                                 & \multicolumn{1}{c|}{NA}                                              & NA                                             & \multicolumn{1}{c|}{5000}                                                             & 5000                                                             \\ \hline
\cite{kang2013mechanism}                                        & CVD                                                                                & \begin{tabular}[c]{@{}c@{}}Back\\ gate\end{tabular}                     & 2/2                                                                                   & \begin{tabular}[c]{@{}c@{}}Thermal\\ ALD\end{tabular}                                     & 150                                                                                 & \multicolumn{1}{c|}{1730}                                            & 161                                            & \multicolumn{1}{c|}{1524}                                                             & 1310                                                             \\ \hline
\cite{kim2018high}                                              & CVD                                                                                & \begin{tabular}[c]{@{}c@{}}Back\\ gate\end{tabular}                            & 30/9                                                                                  & \begin{tabular}[c]{@{}c@{}}Thermal\\ ALD\end{tabular}                                     & 130                                                                                 & \multicolumn{1}{c|}{NA}                                              & NA                                             & \multicolumn{1}{c|}{NA}                                                               & 8600                                                             \\ \hline
\cite{kim2009realization}                                       & \begin{tabular}[c]{@{}c@{}}Mechanical\\ exfoliation\end{tabular}                   & \begin{tabular}[c]{@{}c@{}}Back\\ gate\end{tabular}                     & 2.5/2.5                                                                               & \begin{tabular}[c]{@{}c@{}}Thermal\\ ALD\end{tabular}                                     & 150                                                                                 & \multicolumn{1}{c|}{NA}                                              & NA                                             & \multicolumn{1}{c|}{NA}                                                               & 3000                                                             \\ \hline
\cite{jung2011investigation}                                    & \begin{tabular}[c]{@{}c@{}}Epitaxial\\ grown \\ on SiC\end{tabular}                & \begin{tabular}[c]{@{}c@{}}Back\\ gate\end{tabular}                            & 10/10                                                                                 & \begin{tabular}[c]{@{}c@{}}Anodic\\ Oxidation\end{tabular}                                & NA                                                                                  & \multicolumn{1}{c|}{NA}                                              & NA                                             & \multicolumn{1}{c|}{NA}                                                               & 120                                                              \\ \hline
\textbf{\begin{tabular}[c]{@{}c@{}}Present\\ work\end{tabular}} & \textbf{CVD}                                                                       & \textbf{\begin{tabular}[c]{@{}c@{}}Back\\ gate\end{tabular}}                   & \textbf{5/30}                                                                         & \textbf{\begin{tabular}[c]{@{}c@{}}Sol-gel\\ Alumina\\ +\\ DUV \\ annealing\end{tabular}} & \textbf{70}                                                                         & \multicolumn{1}{c|}{\textbf{2670}}                                   & \textbf{2839}                                  & \multicolumn{1}{c|}{\textbf{1081}}                                                    & \textbf{1319}                                                    \\ \hline
\end{tabular}
\end{table*}

The broadening of the minimum conductivity plateau in transfer characteristics was due to dominant charge impurity scattering from the dielectric \cite{adam2007self}. To support this argument, the XPS in Figure \ref{xps1} (b) shows that there was still OH concentration even after the 20 minutes. These OH present in the oxide could act as charge impurities and scatter the carrier in the graphene channel. These scatters could act as charge impurities at the interface and as scattering centers leading to the reduction in the mobilities \cite{fallahazad2010dielectric}. Table \ref{tab:benchmarktable} shows the summary of passivation layer-based GFETs using different deposition techniques. It can be seen that the present work of GFETs with 0.05M sol-gel alumina-based passivation layer is a low temperature-based process with reasonably good mobility values when compared to others. The process also has 100$\%$ yield with excellent device stability for more than two months.

\section{Conclusion}

The low-cost and low-temperature sol-gel alumina passivation layer was explored with the goal of improving the process yield and device stability of GFET devices. The 0.1 M-based passivation layer led to two problems, (i) process yield was poor as almost half of the devices exhibited crack formation in the channel region, and (ii) the remaining half of working devices showed poor stability, as they degraded in a short time of one week. The degradation has caused a decrease in electron and hole branch of drain current by one order, decreasing mobility by seven times with time respectively. The hypothesis for poor yield and stability was thought to be stress generated during densification annealing of the film. Hence we proposed a film with lower molarity of 0.05 M, which has led to improved process yield with 100 $\%$ working devices without crack, and working devices are also showing good stability even after two months. The devices show no further degradation in drain current, mobility, and trap density. Hence in this work, we proposed an optimized sol-gel alumina passivation layer recipe to achieve the best yield with good stability. The table shows a comparison of our work some of the other reported work. The proposed solution can pave the way to make low-cost and low-temperature-based sol-gel alumina-based GFETs for various applications.

\bibliographystyle{model1-num-names}
\bibliography{sample.bib}

\begin{thebibliography}{45}
\expandafter\ifx\csname natexlab\endcsname\relax\def\natexlab#1{#1}\fi
\providecommand{\bibinfo}[2]{#2}
\ifx\xfnm\relax \def\xfnm[#1]{\unskip,\space#1}\fi
\bibitem[{Novoselov et~al.(2004)Novoselov, Geim, Morozov, Jiang, Zhang,
  Dubonos, Grigorieva, and Firsov}]{novoselov2004electric}
\bibinfo{author}{K.~S. Novoselov}, \bibinfo{author}{A.~K. Geim},
  \bibinfo{author}{S.~V. Morozov}, \bibinfo{author}{D.~Jiang},
  \bibinfo{author}{Y.~Zhang}, \bibinfo{author}{S.~V. Dubonos},
  \bibinfo{author}{I.~V. Grigorieva}, \bibinfo{author}{A.~A. Firsov},
\newblock \bibinfo{title}{Electric field effect in atomically thin carbon
  films},
\newblock \bibinfo{journal}{science} \bibinfo{volume}{306}
  (\bibinfo{year}{2004}) \bibinfo{pages}{666--669}.
\bibitem[{Bolotin et~al.(2008)Bolotin, Sikes, Jiang, Klima, Fudenberg, Hone,
  Kim, and Stormer}]{bolotin2008ultrahigh}
\bibinfo{author}{K.~I. Bolotin}, \bibinfo{author}{K.~Sikes},
  \bibinfo{author}{Z.~Jiang}, \bibinfo{author}{M.~Klima},
  \bibinfo{author}{G.~Fudenberg}, \bibinfo{author}{J.~Hone},
  \bibinfo{author}{P.~Kim}, \bibinfo{author}{H.~Stormer},
\newblock \bibinfo{title}{Ultrahigh electron mobility in suspended graphene},
\newblock \bibinfo{journal}{Solid State Communications} \bibinfo{volume}{146}
  (\bibinfo{year}{2008}) \bibinfo{pages}{351--355}.
\bibitem[{Du et~al.(2008)Du, Skachko, Barker, and Andrei}]{du2008approaching}
\bibinfo{author}{X.~Du}, \bibinfo{author}{I.~Skachko},
  \bibinfo{author}{A.~Barker}, \bibinfo{author}{E.~Y. Andrei},
\newblock \bibinfo{title}{Approaching ballistic transport in suspended
  graphene},
\newblock \bibinfo{journal}{Nature nanotechnology} \bibinfo{volume}{3}
  (\bibinfo{year}{2008}) \bibinfo{pages}{491}.
\bibitem[{Shishir and Ferry(2009)}]{shishir2009velocity}
\bibinfo{author}{R.~Shishir}, \bibinfo{author}{D.~Ferry},
\newblock \bibinfo{title}{Velocity saturation in intrinsic graphene},
\newblock \bibinfo{journal}{Journal of Physics: Condensed Matter}
  \bibinfo{volume}{21} (\bibinfo{year}{2009}) \bibinfo{pages}{344201}.
\bibitem[{Han et~al.(2014)Han, Garcia, Oida, Jenkins, and
  Haensch}]{han2014graphene}
\bibinfo{author}{S.-J. Han}, \bibinfo{author}{A.~V. Garcia},
  \bibinfo{author}{S.~Oida}, \bibinfo{author}{K.~A. Jenkins},
  \bibinfo{author}{W.~Haensch},
\newblock \bibinfo{title}{Graphene radio frequency receiver integrated
  circuit},
\newblock \bibinfo{journal}{Nature communications} \bibinfo{volume}{5}
  (\bibinfo{year}{2014}) \bibinfo{pages}{1--6}.
\bibitem[{Lee et~al.(2012)Lee, Tao, Parrish, Hao, Ruoff, and
  Akinwande}]{lee2012multi}
\bibinfo{author}{J.~Lee}, \bibinfo{author}{L.~Tao}, \bibinfo{author}{K.~N.
  Parrish}, \bibinfo{author}{Y.~Hao}, \bibinfo{author}{R.~S. Ruoff},
  \bibinfo{author}{D.~Akinwande},
\newblock \bibinfo{title}{Multi-finger flexible graphene field effect
  transistors with high bendability},
\newblock \bibinfo{journal}{Applied Physics Letters} \bibinfo{volume}{101}
  (\bibinfo{year}{2012}) \bibinfo{pages}{252109}.
\bibitem[{Park et~al.(2016)Park, Shin, Yogeesh, Lee, Rahimi, and
  Akinwande}]{park2016extremely}
\bibinfo{author}{S.~Park}, \bibinfo{author}{S.~H. Shin}, \bibinfo{author}{M.~N.
  Yogeesh}, \bibinfo{author}{A.~L. Lee}, \bibinfo{author}{S.~Rahimi},
  \bibinfo{author}{D.~Akinwande},
\newblock \bibinfo{title}{Extremely high-frequency flexible graphene thin-film
  transistors},
\newblock \bibinfo{journal}{IEEE Electron Device Letters} \bibinfo{volume}{37}
  (\bibinfo{year}{2016}) \bibinfo{pages}{512--515}.
\bibitem[{Ullah et~al.(2021)Ullah, Yang, Ta, Hasan, Bachmatiuk, Tokarska,
  Trzebicka, Fu, and Rummeli}]{ullah2021graphene}
\bibinfo{author}{S.~Ullah}, \bibinfo{author}{X.~Yang}, \bibinfo{author}{H.~Q.
  Ta}, \bibinfo{author}{M.~Hasan}, \bibinfo{author}{A.~Bachmatiuk},
  \bibinfo{author}{K.~Tokarska}, \bibinfo{author}{B.~Trzebicka},
  \bibinfo{author}{L.~Fu}, \bibinfo{author}{M.~H. Rummeli},
\newblock \bibinfo{title}{Graphene transfer methods: A review},
\newblock \bibinfo{journal}{Nano Research}  (\bibinfo{year}{2021})
  \bibinfo{pages}{1--17}.
\bibitem[{Hu et~al.(2014)Hu, Prasad~Sinha, Ung~Lee, and
  Liehr}]{hu2014substrate}
\bibinfo{author}{Z.~Hu}, \bibinfo{author}{D.~Prasad~Sinha},
  \bibinfo{author}{J.~Ung~Lee}, \bibinfo{author}{M.~Liehr},
\newblock \bibinfo{title}{Substrate dielectric effects on graphene field effect
  transistors},
\newblock \bibinfo{journal}{Journal of Applied Physics} \bibinfo{volume}{115}
  (\bibinfo{year}{2014}) \bibinfo{pages}{194507}.
\bibitem[{Zurutuza{\'a}Elorza et~al.(2015)}]{bzurutuzaaelorza2015highly}
\bibinfo{author}{A.~Zurutuza{\'a}Elorza}, et~al.,
\newblock \bibinfo{title}{Highly air stable passivation of graphene based field
  effect devices},
\newblock \bibinfo{journal}{Nanoscale} \bibinfo{volume}{7}
  (\bibinfo{year}{2015}) \bibinfo{pages}{3558--3564}.
\bibitem[{Alexander-Webber et~al.(2016)Alexander-Webber, Sagade, Aria,
  Van~Veldhoven, Braeuninger-Weimer, Wang, Cabrero-Vilatela, Martin, Sui,
  Connolly et~al.}]{alexander2016encapsulation}
\bibinfo{author}{J.~A. Alexander-Webber}, \bibinfo{author}{A.~A. Sagade},
  \bibinfo{author}{A.~I. Aria}, \bibinfo{author}{Z.~A. Van~Veldhoven},
  \bibinfo{author}{P.~Braeuninger-Weimer}, \bibinfo{author}{R.~Wang},
  \bibinfo{author}{A.~Cabrero-Vilatela}, \bibinfo{author}{M.-B. Martin},
  \bibinfo{author}{J.~Sui}, \bibinfo{author}{M.~R. Connolly}, et~al.,
\newblock \bibinfo{title}{Encapsulation of graphene transistors and vertical
  device integration by interface engineering with atomic layer deposited
  oxide},
\newblock \bibinfo{journal}{2D Materials} \bibinfo{volume}{4}
  (\bibinfo{year}{2016}) \bibinfo{pages}{011008}.
\bibitem[{Chen et~al.(2009)Chen, Xia, Ferry, and Tao}]{chen2009dielectric}
\bibinfo{author}{F.~Chen}, \bibinfo{author}{J.~Xia}, \bibinfo{author}{D.~K.
  Ferry}, \bibinfo{author}{N.~Tao},
\newblock \bibinfo{title}{Dielectric screening enhanced performance in graphene
  fet},
\newblock \bibinfo{journal}{Nano letters} \bibinfo{volume}{9}
  (\bibinfo{year}{2009}) \bibinfo{pages}{2571--2574}.
\bibitem[{Liao and Duan(2010)}]{liao2010graphene}
\bibinfo{author}{L.~Liao}, \bibinfo{author}{X.~Duan},
\newblock \bibinfo{title}{Graphene--dielectric integration for graphene
  transistors},
\newblock \bibinfo{journal}{Materials Science and Engineering: R: Reports}
  \bibinfo{volume}{70} (\bibinfo{year}{2010}) \bibinfo{pages}{354--370}.
\bibitem[{Wang et~al.(2018)Wang, Huang, Chi, Al-Hashimi, Marks, and
  Facchetti}]{wang2018high}
\bibinfo{author}{B.~Wang}, \bibinfo{author}{W.~Huang},
  \bibinfo{author}{L.~Chi}, \bibinfo{author}{M.~Al-Hashimi},
  \bibinfo{author}{T.~J. Marks}, \bibinfo{author}{A.~Facchetti},
\newblock \bibinfo{title}{High-k gate dielectrics for emerging flexible and
  stretchable electronics},
\newblock \bibinfo{journal}{Chemical reviews} \bibinfo{volume}{118}
  (\bibinfo{year}{2018}) \bibinfo{pages}{5690--5754}.
\bibitem[{Giubileo and Di~Bartolomeo(2017)}]{giubileo2017role}
\bibinfo{author}{F.~Giubileo}, \bibinfo{author}{A.~Di~Bartolomeo},
\newblock \bibinfo{title}{The role of contact resistance in graphene
  field-effect devices},
\newblock \bibinfo{journal}{Progress in Surface Science} \bibinfo{volume}{92}
  (\bibinfo{year}{2017}) \bibinfo{pages}{143--175}.
\bibitem[{Neumaier et~al.(2019)Neumaier, Pindl, and
  Lemme}]{neumaier2019integrating}
\bibinfo{author}{D.~Neumaier}, \bibinfo{author}{S.~Pindl},
  \bibinfo{author}{M.~C. Lemme},
\newblock \bibinfo{title}{Integrating graphene into semiconductor fabrication
  lines},
\newblock \bibinfo{journal}{Nature materials} \bibinfo{volume}{18}
  (\bibinfo{year}{2019}) \bibinfo{pages}{525--529}.
\bibitem[{Jee et~al.(2009)Jee, Han, Hwang, Kim, Kim, Kim, and
  Hwang}]{jee2009pentacene}
\bibinfo{author}{H.-g. Jee}, \bibinfo{author}{J.-H. Han},
  \bibinfo{author}{H.-N. Hwang}, \bibinfo{author}{B.~Kim},
  \bibinfo{author}{H.-s. Kim}, \bibinfo{author}{Y.~D. Kim},
  \bibinfo{author}{C.-C. Hwang},
\newblock \bibinfo{title}{Pentacene as protection layers of graphene on sic
  surfaces},
\newblock \bibinfo{journal}{Applied Physics Letters} \bibinfo{volume}{95}
  (\bibinfo{year}{2009}) \bibinfo{pages}{093107}.
\bibitem[{Long et~al.(2012)Long, Manning, Burke, Szafranek, Visimberga,
  Thompson, Greer, Povey, MacHale, Lejosne et~al.}]{long2012non}
\bibinfo{author}{B.~Long}, \bibinfo{author}{M.~Manning},
  \bibinfo{author}{M.~Burke}, \bibinfo{author}{B.~N. Szafranek},
  \bibinfo{author}{G.~Visimberga}, \bibinfo{author}{D.~Thompson},
  \bibinfo{author}{J.~C. Greer}, \bibinfo{author}{I.~M. Povey},
  \bibinfo{author}{J.~MacHale}, \bibinfo{author}{G.~Lejosne}, et~al.,
\newblock \bibinfo{title}{Non-covalent functionalization of graphene using
  self-assembly of alkane-amines},
\newblock \bibinfo{journal}{Advanced Functional Materials} \bibinfo{volume}{22}
  (\bibinfo{year}{2012}) \bibinfo{pages}{717--725}.
\bibitem[{Mayorov et~al.(2011)Mayorov, Gorbachev, Morozov, Britnell, Jalil,
  Ponomarenko, Blake, Novoselov, Watanabe, Taniguchi
  et~al.}]{mayorov2011micrometer}
\bibinfo{author}{A.~S. Mayorov}, \bibinfo{author}{R.~V. Gorbachev},
  \bibinfo{author}{S.~V. Morozov}, \bibinfo{author}{L.~Britnell},
  \bibinfo{author}{R.~Jalil}, \bibinfo{author}{L.~A. Ponomarenko},
  \bibinfo{author}{P.~Blake}, \bibinfo{author}{K.~S. Novoselov},
  \bibinfo{author}{K.~Watanabe}, \bibinfo{author}{T.~Taniguchi}, et~al.,
\newblock \bibinfo{title}{Micrometer-scale ballistic transport in encapsulated
  graphene at room temperature},
\newblock \bibinfo{journal}{Nano letters} \bibinfo{volume}{11}
  (\bibinfo{year}{2011}) \bibinfo{pages}{2396--2399}.
\bibitem[{Kim et~al.(2009)Kim, Nah, Jo, Shahrjerdi, Colombo, Yao, Tutuc, and
  Banerjee}]{kim2009realization}
\bibinfo{author}{S.~Kim}, \bibinfo{author}{J.~Nah}, \bibinfo{author}{I.~Jo},
  \bibinfo{author}{D.~Shahrjerdi}, \bibinfo{author}{L.~Colombo},
  \bibinfo{author}{Z.~Yao}, \bibinfo{author}{E.~Tutuc}, \bibinfo{author}{S.~K.
  Banerjee},
\newblock \bibinfo{title}{Realization of a high mobility dual-gated graphene
  field-effect transistor with al 2 o 3 dielectric},
\newblock \bibinfo{journal}{Applied Physics Letters} \bibinfo{volume}{94}
  (\bibinfo{year}{2009}) \bibinfo{pages}{062107}.
\bibitem[{Kang et~al.(2013)Kang, Lee, Lee, Park, Cho, Lim, Hwang, and
  Lee}]{kang2013mechanism}
\bibinfo{author}{C.~G. Kang}, \bibinfo{author}{Y.~G. Lee},
  \bibinfo{author}{S.~K. Lee}, \bibinfo{author}{E.~Park},
  \bibinfo{author}{C.~Cho}, \bibinfo{author}{S.~K. Lim}, \bibinfo{author}{H.~J.
  Hwang}, \bibinfo{author}{B.~H. Lee},
\newblock \bibinfo{title}{Mechanism of the effects of low temperature al 2 o 3
  passivation on graphene field effect transistors},
\newblock \bibinfo{journal}{Carbon} \bibinfo{volume}{53} (\bibinfo{year}{2013})
  \bibinfo{pages}{182--187}.
\bibitem[{Vervuurt et~al.(2017)Vervuurt, Karasulu, Verheijen, Kessels, and
  Bol}]{vervuurt2017uniform}
\bibinfo{author}{R.~H. Vervuurt}, \bibinfo{author}{B.~Karasulu},
  \bibinfo{author}{M.~A. Verheijen}, \bibinfo{author}{W.~E.~M. Kessels},
  \bibinfo{author}{A.~A. Bol},
\newblock \bibinfo{title}{Uniform atomic layer deposition of al2o3 on graphene
  by reversible hydrogen plasma functionalization},
\newblock \bibinfo{journal}{Chemistry of Materials} \bibinfo{volume}{29}
  (\bibinfo{year}{2017}) \bibinfo{pages}{2090--2100}.
\bibitem[{Kim et~al.(2018)Kim, Kim, Heo, Lee, Lee, Chang, and
  Lee}]{kim2018high}
\bibinfo{author}{Y.~J. Kim}, \bibinfo{author}{S.~M. Kim},
  \bibinfo{author}{S.~Heo}, \bibinfo{author}{H.~Lee}, \bibinfo{author}{H.~I.
  Lee}, \bibinfo{author}{K.~E. Chang}, \bibinfo{author}{B.~H. Lee},
\newblock \bibinfo{title}{High-pressure oxygen annealing of al2o3 passivation
  layer for performance enhancement of graphene field-effect transistors},
\newblock \bibinfo{journal}{Nanotechnology} \bibinfo{volume}{29}
  (\bibinfo{year}{2018}) \bibinfo{pages}{055202}.
\bibitem[{Brinker and Scherer(2013)}]{brinker2013sol}
\bibinfo{author}{C.~J. Brinker}, \bibinfo{author}{G.~W. Scherer},
  \bibinfo{title}{Sol-gel science: the physics and chemistry of sol-gel
  processing}, \bibinfo{publisher}{Academic press}, \bibinfo{year}{2013}.
\bibitem[{Cochran et~al.(2019)Cochran, Woods, Johnson, Page, and
  Boettcher}]{cochran2019unique}
\bibinfo{author}{E.~A. Cochran}, \bibinfo{author}{K.~N. Woods},
  \bibinfo{author}{D.~W. Johnson}, \bibinfo{author}{C.~J. Page},
  \bibinfo{author}{S.~W. Boettcher},
\newblock \bibinfo{title}{Unique chemistries of metal-nitrate precursors to
  form metal-oxide thin films from solution: materials for electronic and
  energy applications},
\newblock \bibinfo{journal}{Journal of Materials Chemistry A}
  \bibinfo{volume}{7} (\bibinfo{year}{2019}) \bibinfo{pages}{24124--24149}.
\bibitem[{Park et~al.(2017)Park, Kim, Lee, Sung, and Yoon}]{park2017sol}
\bibinfo{author}{S.~Park}, \bibinfo{author}{C.-H. Kim}, \bibinfo{author}{W.-J.
  Lee}, \bibinfo{author}{S.~Sung}, \bibinfo{author}{M.-H. Yoon},
\newblock \bibinfo{title}{Sol-gel metal oxide dielectrics for
  all-solution-processed electronics},
\newblock \bibinfo{journal}{Materials Science and Engineering: R: Reports}
  \bibinfo{volume}{114} (\bibinfo{year}{2017}) \bibinfo{pages}{1--22}.
\bibitem[{Khatibani and Rozati(2014)}]{khatibani2014growth}
\bibinfo{author}{A.~B. Khatibani}, \bibinfo{author}{S.~Rozati},
\newblock \bibinfo{title}{Growth and molarity effects on properties of alumina
  thin films obtained by spray pyrolysis},
\newblock \bibinfo{journal}{Materials science in semiconductor processing}
  \bibinfo{volume}{18} (\bibinfo{year}{2014}) \bibinfo{pages}{80--87}.
\bibitem[{Kozuka et~al.(2003)Kozuka, Takenaka, Tokita, Hirano, Higashi, and
  Hamatani}]{kozuka2003stress}
\bibinfo{author}{H.~Kozuka}, \bibinfo{author}{S.~Takenaka},
  \bibinfo{author}{H.~Tokita}, \bibinfo{author}{T.~Hirano},
  \bibinfo{author}{Y.~Higashi}, \bibinfo{author}{T.~Hamatani},
\newblock \bibinfo{title}{Stress and cracks in gel-derived ceramic coatings and
  thick film formation},
\newblock \bibinfo{journal}{Journal of Sol-Gel Science and Technology}
  \bibinfo{volume}{26} (\bibinfo{year}{2003}) \bibinfo{pages}{681--686}.
\bibitem[{Liu et~al.(2018)Liu, Zhu, Sun, Xu, and Noh}]{liu2018solution}
\bibinfo{author}{A.~Liu}, \bibinfo{author}{H.~Zhu}, \bibinfo{author}{H.~Sun},
  \bibinfo{author}{Y.~Xu}, \bibinfo{author}{Y.-Y. Noh},
\newblock \bibinfo{title}{Solution processed metal oxide high-$\kappa$
  dielectrics for emerging transistors and circuits},
\newblock \bibinfo{journal}{Advanced Materials} \bibinfo{volume}{30}
  (\bibinfo{year}{2018}) \bibinfo{pages}{1706364}.
\bibitem[{Park et~al.(2015)Park, Kim, Jo, Sung, Kim, Lee, Kim, Kim, Yi, Kim
  et~al.}]{park2015depth}
\bibinfo{author}{S.~Park}, \bibinfo{author}{K.-H. Kim}, \bibinfo{author}{J.-W.
  Jo}, \bibinfo{author}{S.~Sung}, \bibinfo{author}{K.-T. Kim},
  \bibinfo{author}{W.-J. Lee}, \bibinfo{author}{J.~Kim}, \bibinfo{author}{H.~J.
  Kim}, \bibinfo{author}{G.-R. Yi}, \bibinfo{author}{Y.-H. Kim}, et~al.,
\newblock \bibinfo{title}{In-depth studies on rapid photochemical activation of
  various sol--gel metal oxide films for flexible transparent electronics},
\newblock \bibinfo{journal}{Advanced Functional Materials} \bibinfo{volume}{25}
  (\bibinfo{year}{2015}) \bibinfo{pages}{2807--2815}.
\bibitem[{Kalaivani and Kottantharayil(2015)}]{kalaivani2015spray}
\bibinfo{author}{S.~Kalaivani}, \bibinfo{author}{A.~Kottantharayil},
\newblock \bibinfo{title}{Spray coated aluminum oxide thin film for p-type
  crystalline silicon surface passivation},
\newblock in: \bibinfo{booktitle}{Photovoltaic Specialist Conference (PVSC),
  2015 IEEE 42nd}, \bibinfo{organization}{IEEE}, pp. \bibinfo{pages}{1--4}.
\bibitem[{Park et~al.(2016)Park, Kim, Fukidome, Suemitsu, Otsuji, Cho, and
  Suemitsu}]{park2016solution}
\bibinfo{author}{G.-H. Park}, \bibinfo{author}{K.-S. Kim},
  \bibinfo{author}{H.~Fukidome}, \bibinfo{author}{T.~Suemitsu},
  \bibinfo{author}{T.~Otsuji}, \bibinfo{author}{W.-J. Cho},
  \bibinfo{author}{M.~Suemitsu},
\newblock \bibinfo{title}{Solution-processed al2o3 gate dielectrics for
  graphene field-effect transistors},
\newblock \bibinfo{journal}{Japanese Journal of Applied Physics}
  \bibinfo{volume}{55} (\bibinfo{year}{2016}) \bibinfo{pages}{091502}.
\bibitem[{Kim et~al.(2019)Kim, Fukidome, and Suemitsu}]{kim2019direct}
\bibinfo{author}{K.-S. Kim}, \bibinfo{author}{H.~Fukidome},
  \bibinfo{author}{M.~Suemitsu},
\newblock \bibinfo{title}{Direct formation of solution-based al2o3 on epitaxial
  graphene surface for sensor applications},
\newblock \bibinfo{journal}{Sensors and Materials} \bibinfo{volume}{31}
  (\bibinfo{year}{2019}) \bibinfo{pages}{2291--2301}.
\bibitem[{Bae et~al.(2014)Bae, Kim, Jung, and Cho}]{bae2014fabrication}
\bibinfo{author}{T.-E. Bae}, \bibinfo{author}{H.~Kim},
  \bibinfo{author}{J.~Jung}, \bibinfo{author}{W.-J. Cho},
\newblock \bibinfo{title}{Fabrication of high-performance graphene field-effect
  transistor with solution-processed al2o3 sensing membrane},
\newblock \bibinfo{journal}{Applied Physics Letters} \bibinfo{volume}{104}
  (\bibinfo{year}{2014}) \bibinfo{pages}{153506}.
\bibitem[{Van~de Leest(1995)}]{van1995uv}
\bibinfo{author}{R.~Van~de Leest},
\newblock \bibinfo{title}{Uv photo-annealing of thin sol-gel films},
\newblock \bibinfo{journal}{Applied surface science} \bibinfo{volume}{86}
  (\bibinfo{year}{1995}) \bibinfo{pages}{278--285}.
\bibitem[{Zhong et~al.(2015)Zhong, Zhang, Xu, Qiu, and
  Peng}]{zhong2015comparison}
\bibinfo{author}{H.~Zhong}, \bibinfo{author}{Z.~Zhang},
  \bibinfo{author}{H.~Xu}, \bibinfo{author}{C.~Qiu}, \bibinfo{author}{L.-M.
  Peng},
\newblock \bibinfo{title}{Comparison of mobility extraction methods based on
  field-effect measurements for graphene},
\newblock \bibinfo{journal}{AIP Advances} \bibinfo{volume}{5}
  (\bibinfo{year}{2015}) \bibinfo{pages}{057136}.
\bibitem[{Sul et~al.(2016)Sul, Kim, Choi, Kil, Park, and
  Lee}]{sul2016reduction}
\bibinfo{author}{O.~Sul}, \bibinfo{author}{K.~Kim}, \bibinfo{author}{E.~Choi},
  \bibinfo{author}{J.~Kil}, \bibinfo{author}{W.~Park}, \bibinfo{author}{S.-B.
  Lee},
\newblock \bibinfo{title}{Reduction of hole doping of chemical vapor deposition
  grown graphene by photoresist selection and thermal treatment},
\newblock \bibinfo{journal}{Nanotechnology} \bibinfo{volume}{27}
  (\bibinfo{year}{2016}) \bibinfo{pages}{505205}.
\bibitem[{Nagashio et~al.(2009)Nagashio, Nishimura, Kita, and
  Toriumi}]{nagashio2009metal}
\bibinfo{author}{K.~Nagashio}, \bibinfo{author}{T.~Nishimura},
  \bibinfo{author}{K.~Kita}, \bibinfo{author}{A.~Toriumi},
\newblock \bibinfo{title}{Metal/graphene contact as a performance killer of
  ultra-high mobility graphene analysis of intrinsic mobility and contact
  resistance},
\newblock in: \bibinfo{booktitle}{2009 IEEE International Electron Devices
  Meeting (IEDM)}, \bibinfo{organization}{IEEE}, pp. \bibinfo{pages}{1--4}.
\bibitem[{Chen et~al.(2008)Chen, Jang, Adam, Fuhrer, Williams, and
  Ishigami}]{chen2008charged}
\bibinfo{author}{J.-H. Chen}, \bibinfo{author}{C.~Jang},
  \bibinfo{author}{S.~Adam}, \bibinfo{author}{M.~Fuhrer},
  \bibinfo{author}{E.~D. Williams}, \bibinfo{author}{M.~Ishigami},
\newblock \bibinfo{title}{Charged-impurity scattering in graphene},
\newblock \bibinfo{journal}{Nature physics} \bibinfo{volume}{4}
  (\bibinfo{year}{2008}) \bibinfo{pages}{377}.
\bibitem[{Katoch and Ishigami(2012)}]{katoch2012impact}
\bibinfo{author}{J.~Katoch}, \bibinfo{author}{M.~Ishigami},
\newblock \bibinfo{title}{Impact of calcium on transport property of graphene},
\newblock \bibinfo{journal}{Solid state communications} \bibinfo{volume}{152}
  (\bibinfo{year}{2012}) \bibinfo{pages}{60--63}.
\bibitem[{Adam et~al.(2007)Adam, Hwang, Galitski, and Sarma}]{adam2007self}
\bibinfo{author}{S.~Adam}, \bibinfo{author}{E.~Hwang},
  \bibinfo{author}{V.~Galitski}, \bibinfo{author}{S.~D. Sarma},
\newblock \bibinfo{title}{A self-consistent theory for graphene transport},
\newblock \bibinfo{journal}{Proceedings of the National Academy of Sciences}
  \bibinfo{volume}{104} (\bibinfo{year}{2007}) \bibinfo{pages}{18392--18397}.
\bibitem[{Wang et~al.(2010)Wang, Wu, Cong, Shang, and Yu}]{wang2010hysteresis}
\bibinfo{author}{H.~Wang}, \bibinfo{author}{Y.~Wu}, \bibinfo{author}{C.~Cong},
  \bibinfo{author}{J.~Shang}, \bibinfo{author}{T.~Yu},
\newblock \bibinfo{title}{Hysteresis of electronic transport in graphene
  transistors},
\newblock \bibinfo{journal}{ACS nano} \bibinfo{volume}{4}
  (\bibinfo{year}{2010}) \bibinfo{pages}{7221--7228}.
\bibitem[{Lee et~al.(2013)Lee, Kang, Cho, Kim, Hwang, and
  Lee}]{lee2013quantitative}
\bibinfo{author}{Y.~G. Lee}, \bibinfo{author}{C.~G. Kang},
  \bibinfo{author}{C.~Cho}, \bibinfo{author}{Y.~Kim}, \bibinfo{author}{H.~J.
  Hwang}, \bibinfo{author}{B.~H. Lee},
\newblock \bibinfo{title}{Quantitative analysis of hysteretic reactions at the
  interface of graphene and sio2 using the short pulse i--v method},
\newblock \bibinfo{journal}{Carbon} \bibinfo{volume}{60} (\bibinfo{year}{2013})
  \bibinfo{pages}{453--460}.
\bibitem[{Jung et~al.(2011)Jung, Handa, Takahashi, Fukidome, Suemitsu, Otsuji,
  and Suemitsu}]{jung2011investigation}
\bibinfo{author}{M.-H. Jung}, \bibinfo{author}{H.~Handa},
  \bibinfo{author}{R.~Takahashi}, \bibinfo{author}{H.~Fukidome},
  \bibinfo{author}{T.~Suemitsu}, \bibinfo{author}{T.~Otsuji},
  \bibinfo{author}{M.~Suemitsu},
\newblock \bibinfo{title}{Investigation of graphene field effect transistors
  with al2o3 gate dielectrics formed by metal oxidation},
\newblock \bibinfo{journal}{Japanese Journal of Applied Physics}
  \bibinfo{volume}{50} (\bibinfo{year}{2011}) \bibinfo{pages}{070111}.
\bibitem[{Fallahazad et~al.(2010)Fallahazad, Kim, Colombo, and
  Tutuc}]{fallahazad2010dielectric}
\bibinfo{author}{B.~Fallahazad}, \bibinfo{author}{S.~Kim},
  \bibinfo{author}{L.~Colombo}, \bibinfo{author}{E.~Tutuc},
\newblock \bibinfo{title}{Dielectric thickness dependence of carrier mobility
  in graphene with hfo 2 top dielectric},
\newblock \bibinfo{journal}{Applied Physics Letters} \bibinfo{volume}{97}
  (\bibinfo{year}{2010}) \bibinfo{pages}{123105}.

\end{thebibliography}







\end{document}